\documentclass[draft]{agujournal2019}
\usepackage{url} %this package should fix any errors with URLs

\usepackage{times}
\usepackage[final]{editing}
\usepackage{soul}

\draftfalse
\journalname{Geophysical Research Letters}
\graphicspath{{./Figures/}{./Figures/F2/}{./Figures/F1/}}

\usepackage[T1]{fontenc}

\begin{document}
%TC:ignore
\title{Dust storm-enhanced gravity wave activity in the Martian thermosphere observed by MAVEN and implication for atmospheric escape}
\authors{Erdal Yi\u git\affil{1}, Alexander S. Medvedev\affil{2}, Mehdi Benna\affil{3,4}, Bruce M. Jakosky\affil{5}}
\affiliation{1}{George Mason University, Department of Physics and Astronomy.}
\affiliation{2}{Max Planck Institute for Solar System Research, G\o"ttingen, Germany.}
\affiliation{3}{University of Maryland Baltimore County, Baltimore, MD, USA.}
\affiliation{4}{Solar System Exploration Division, NASA Goddard Space Flight Center, Greenbelt, MD, USA.}
\affiliation{5}{Laboratory for Atmospheric and Space Physics,  University of Colorado, CO, USA.}
\correspondingauthor{Erdal Yi\u git}{eyigit@gmu.edu}
%  Each must be 140 characters
\begin{keypoints}
\item Thermospheric gravity wave activity doubles during the dust storm.
\item Gravity wave induced density fluctuations in the thermosphere are up to 40\% during the peak storm phase.
\item Gravity waves significantly increase Hydrogen escape flux by modulating temperature fluctuations.

\end{keypoints}
%TC:endignore
\begin{abstract}
Lower atmospheric global dust storms affect the small- and large-scale weather and variability of the whole Martian atmosphere. Analysis of the CO$_2$ density data from the Neutral Gas and Ion Mass Spectrometer instrument (NGIMS) on board NASA's Mars Atmosphere Volatile EvolutioN (MAVEN) spacecraft show a remarkable increase of GW-induced density fluctuations in the thermosphere during the 2018 major dust storm with distinct latitude and local time variability. The mean thermospheric GW activity increases by a factor of two during the storm event. The magnitude of relative density perturbations is around 20\% on average and 40\% locally. One and a half months later, the GW activity gradually decreases. Enhanced temperature disturbances in the Martian thermosphere can facilitate atmospheric escape. For the first time, we estimate that, for a 20\% and 40\% GW-induced disturbances, the net increase of Jeans escape flux of hydrogen is a factor of 1.3 and 2, respectively.
\end{abstract}

\section*{Plain Language Summary}
Atmospheric gravity waves play an important dynamical and thermodynamical role in coupling the different atmospheric layers, especially on Earth and Mars. We study the effects of a planet-encircling major dust storm on thermospheric gravity wave activity and estimate for the first time a potential influence of gravity waves on atmospheric escape on Mars. Gravity activity measured in terms of relative density fluctuations increases by a factor of two during the peak phase of the storm. We show that larger-amplitude gravity waves facilitate atmospheric escape of hydrogen from Mars' upper atmosphere. For 40\% gravity wave-induced relative disturbances of temperature, the net escape rate doubles. 

\section{Introduction}
\label{sec:intro}
Dust greatly impacts the dynamics and thermodynamics of the entire \addE{entire} Martian atmosphere \cite{Haberle_etal82, ZurekMartin93, Bell_etal07, Cantor07, Clancy_etal10a, Heavens_etal11, Medvedev_etal13, Jain_etal20, Wu_etal20, Liuzzi_etal20}. During storms, regolith particles are raised from the surface and modify temperature by absorbing more solar radiation within the atmosphere and obstructing heating of the lowermost layers \cite{Gierasch.Goody72, Rafkin09}. Once dust is airborne, sedimentation may take up to several months. Depending on the scale, storms can be regional or global with wide-reaching implications for the planetary climate.

Dust storms affect circulation at all scales, in particular the atmospheric gravity wave (GW) activity. GWs (or buoyancy waves) are ubiquitous features of all planetary atmospheres \cite<e.g., see recent reviews of>{YigitMedvedev19, MedvedevYigit19}. They have been extensively studied on Earth since the 1960s, when their essential role in coupling atmospheric layers was recognized. On Mars, GWs have been observed by a number of satellites \cite{Fritts_etal06, Tolson_etal07, Yigit_etal15b, England_etal17, Jesch_etal19, Vals_etal19, Siddle_etal19} and studied with numerical models \cite{Parish_etal09, Medvedev_etal13, Walterscheid_etal13, Imamura_etal16, Yigit_etal18, Kuroda_etal19}. The main mechanism by which GWs affect the dynamics and state of the atmosphere is transporting energy and momentum from denser lower levels and depositing them in the thinner upper atmosphere. The latter is also the region where atmospheric escape takes place \cite{Walterscheid_etal13, Chaffin_etal18}, however the impact of GWs on the escape rate has not been considered \addE{observationally} before, to the best of our knowledge.

Thermospheric response to global dust storms (GDS) have been extensively studied during the major event of 2018. In particular, \citeA{Jain_etal20} and \citeA{Elrod_etal20} characterized large-scale thermospheric effects of the GDS. Recently, based on the Ar measurements with the Neutral Gas and Ion Spectrometer (NGIMS) instrument on board the Mars Atmosphere Volatile Evolution Mission (MAVEN) orbiter, \citeA{Leelavathi_etal20} reported on the increase of GW activity during the storm of 2018 in the thermosphere. In our paper, we also quantify thermospheric GW activity during different phases of the planet encircling dust storm that commenced on 1 June 2018 using NGIMS' measurements of CO$_2$ and discuss possible implications for atmospheric escape.

\section{Materials and Methods}
\label{sec:methods}

%%% Data coverage
\subsection{Data Sets Analyzed}
\label{sec:data-coverage-my}

For the analysis of the GW activity before and during the planet-encircling global dust storm, we consider data from the NGIMS instrument onboard the MAVEN spacecraft from 1 May 2018 till 30 September 2018, corresponding to $L_s=167^\circ-259.6^\circ$ in Martian Year (MY) 34. In the analysis to be presented we also compare the dust-storm GW activity in MY34 with a low-dust period one Martian year earlier. For this,  the low-dust period in MY 33 with solar longitudes $L_s=171.7^\circ-191.6^\circ$ (20 June-25 July 2016) is compared with a representative dust storm period in MY 34, $L_s=202.8^\circ-224.2^\circ$ (1 July-5 August 2018), when MAVEN had comparable latitude and local time coverage. The details of the data used and orbital coverage are provided in the supporting information and figures.

%%% GW analysis %%%
\subsection{Observational Analysis of Wave Activity}
\label{sec:observ-analys-wave}
Calculation of the GW fluctuations requires information on the background field. For this, we use a 7-th order polynomial fit to the logarithm of the CO$_2$ (carbon dioxide) density profiles to determine the mean field. Polynomial fit technique has been used in a number of previous studies of GWs on Mars \cite{Yigit_etal15b,England_etal17,Siddle_etal19} and Earth \cite{Randall_etal17}. In order to calculate the GW-induced fluctuations we subtract the background mean density (i.e., the polynomial fit) from the instantaneous measurements to determine the GW disturbance as:
\begin{equation}
  \label{eq:1}
  \rho^\prime = \rho-\bar{\rho},
\end{equation}
where the $\bar{\rho}$ is the background (polynomial fitted density) and $\rho$ is the measured (instantaneous) CO$_2$ density. The relative density perturbation in percentage is then given by dividing the density fluctuations by the background mean\addE{, $\rho^\prime / \bar{\rho}$}, and multiplying by 100. This analysis is used for each orbit, including the inbound and the outbound pass. %An illustration of the analysis is given in ``Example of GW analysis" below.
 
%\textit{Data binning:}
In order to evaluate the variation of the GW activity for the period of one month before the onset of the storm to the end phases (1 May 2018 till 30 September 2018), we first organize all 683 orbits in $\sim $15-day ($\sim $15-sol) intervals. Then 15-day mean GW-induced \addE{relative} density fluctuations\addE{, $\overline{\rho^\prime/\bar{\rho}}$,} are calculated from the average of data points within each bin as a function of altitude, longitude, latitude, and local time as presented in Figure 1, using 5 km, $30^\circ$, $5^\circ$, and 1 hour bins, respectively. For the comparison of MY34 dust-storm period (1 July-5 August 2018) to MY33 low-dust period (20 June-25 July 2016) presented in Figure 2, we focused on the data between 160 and 200 km, and binned them in terms of 5 km, $20^\circ$ longitude, $5^\circ$ latitude, and 0.5 hour bins. This for example means that data point at the altitude level 160 km represents the average value for the bin from 160-165 km. \addE{The typical uncertainty in the mean GW activity is about $0.26-0.7\%$.}
\addE{The details of the GW analysis and the uncertainty are discussed in the supporting material and figures.}

\section{Results}

\begin{figure}
  \vspace{-1.8cm}
  \hspace*{-1.cm}
  \includegraphics[width=1.1\textwidth]{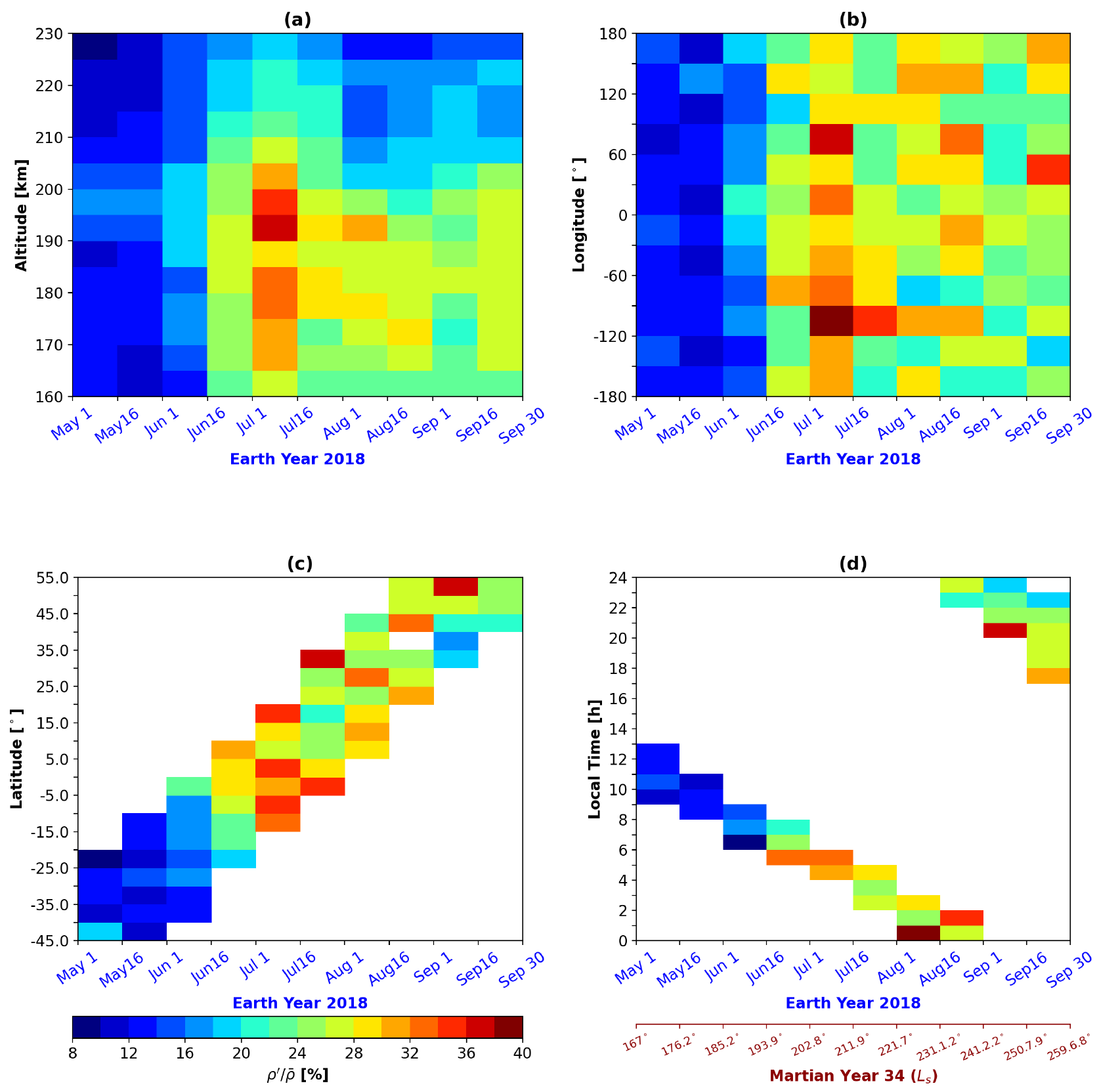}
  \caption{(a) Altitude, (b) longitude, (c) latitude, and (d) local time variations of the gravity wave activity in terms of relative carbon dioxide density perturbations $\rho^\prime/\bar{\rho}$ before and during the different phases of the dust storm in MY=34 from solar longitudes $L_s =167^\circ-259^\circ$ (1 May-30 September 2018). All data are averaged over a $\sim$15-day time intervals. Data binning is performed in terms of 5 km, $30^\circ$,$5^\circ$, and 1 hour bins in (a)-(d), respectively.}
  \label{fig:binning}
\end{figure}

Variations of the GW-induced CO$_2$  relative density fluctuations before and during the storm are shown in Figure \ref{fig:binning}. The average fluctuations increase from 8-12\% before the onset (1 June 2018) and rapidly increase afterwards, peaking with $\sim$40\% between 1-16 July ($L_s=202.8^\circ- 211.9^\circ$) around 190-195 km. The GW activity increases at all thermospheric heights (panel a), but the maximum occurs between $\sim$165-205 km. Panels (b-d) show the longitude, latitude, and local time variations of GW activity during the same period, focusing on the region between 165-185 km. Enhanced activity is systematically seen there in all analyses. During this period, MAVEN's observations sampled low latitudes ($15^\circ S-20^\circ$N)  and local nighttime (4-6 h). They demonstrate some difference in GW activity with larger values in the low-latitude southern (spring) hemisphere than the low-latitude northern hemisphere. MAVEN's orbit and coverage change in latitude and local time over the analyzed period (see supplementary Figure S1). From the pre-storm period toward the peak of the GDS, the spacecraft coverage moves from southern midlatitudes ($45^\circ $S$-25^\circ$S) to equatorial latitudes ($15^\circ $S$-20^\circ$N) and from local times 9-13 h to 4-6 h. These changes should be accounted for in order to isolate them from dust-induced effects.

For that, we consistently compared the GW activity during the 2018 GDS against measurements for low-dust conditions one Martian year earlier. MAVEN's coverage changed with $L_s$, latitude and local time due to specifics of the orbit.  We identified two periods with similar seasonal and spatial orbit characteristics: 20 June-25 July 2016 (MY 33, $L_s=171.7^\circ-191.6^\circ$) and 1 July-6 August 2018 (MY 34, $L_s=202.8^\circ-224.2^\circ$) (see supplementary information). Figure \ref{fig:2016vs2018} shows the altitude, longitude, latitude and local time variations of GW activity during these two periods. Similar to Figure \ref{fig:binning}, averaging over the height interval 165-185 km has been performed. It is seen that GW activity is about two times larger during the storm. The southern hemisphere (SH) values are larger than those in the northern hemisphere (NH) for both the low-dust and dusty conditions. Figure \ref{fig:2016vs2018b} shows another perspective of how GW activity increases as a consequence of the GDS, presented in terms of global distributions of wave-induced density fluctuations during the chosen periods. Here, we binned the nighttime (local times 1.5-4.5 h) data between 165-185 km in terms of latitude and longitude. The effect of the storm on the GWs is remarkable: the activity is around 8-10\% under low-dust conditions and increases to more than 20\% globally and even 40\% locally.

\section{Discussion}
\label{sec:discussion}

\subsection{Mechanism of Dust-Induced Gravity Wave Enhancement}
The observed enhancement of GW activity in the upper atmosphere during the dust storm agrees well with the results of \citeA{Leelavathi_etal20}, but is quite unexpected. \addE{Since the gravity wave energy and momentum flux are proportional to the square of the wave amplitude, the increase in observed amplitude is, in fact, even higher in terms of these dynamically important quantities.} An overall effect of storms on the lower atmosphere is the convective \cite<Figure 1c of>{Kuroda_etal20} and baroclinic \cite<Figure 2 of>{Kuroda_etal07} stabilization of the circulation: smaller lapse rates impede development of convection, and intensified zonal jets inhibit formation of larger-scale weather disturbances. This effectively suppresses the major mechanisms of GW generation in the lower atmosphere. Observations by Mars Climate Sounder provided evidence of a reduction of GW activity below 30 km by several times during the 2018 GDS \cite{Heavens_etal20a}. Airborne aerosol particles do not rise above $\sim$70 km. Why does the wave activity in the upper atmosphere increase then?

In the absence of other indications favoring in-situ wave generation, a plausible explanation is related to changes in the upward propagation of GWs. The latter primarily depends on the background winds and wave dissipation, such as nonlinear breaking and molecular diffusion \cite{HickeyCole88, Yigit_etal08, Parish_etal09, Hickey_etal15}. GW harmonics are absorbed by the mean flow, when their horizontal phase velocity approaches the ambient wind speed. Large local vertical gradients within a wave make harmonics prone to break-down and/or enhanced dissipation. During dust storms, the middle atmosphere circulation undergoes substantial changes due to the storm-induced radiative heating, which in turn modulate upward propagation and dissipation of GWs. The observed increase in thermospheric GW activity indicates that GW harmonics encounter more favorable propagation conditions during the dust storm. High-resolution simulations have demonstrated that the middle atmospheric GW activity increases despite the reduction in the lower atmosphere \cite{Kuroda_etal20}, thus supporting this hypothesis. Although the details of this mechanism are not fully understood, it provides evidence for yet another consequence of Martian dust storms: they facilitate vertical coupling between atmospheric layers.

The increase of GW activity is even more unexpected in view of the recent finding that wave amplitudes observed by NGIMS typically decrease in proportion to the upper thermospheric temperature \cite{England_etal17, Terada_etal17, Vals_etal19}. The mechanism that likely controls such behavior is wave saturation due to convective instability, which permits larger amplitudes when the atmosphere is colder. However, the thermosphere warms during the 2018 dust storm event \cite{Jain_etal20}, which would imply weaker GW activity contrary to our results. 

\subsection{Gravity Waves and Atmospheric Escape}

The observed enhancement of GW activity in the upper atmosphere during the MY34 GDS has far-reaching implications for the state as well as short- and long-term evolution of the Martian atmosphere. Recent ExoMars Trace Gas Orbiter observations reported a sudden increase of water vapor in the middle atmosphere during the storm, which was delivered there from below by the thermally-enhanced meridional circulation \cite{Vandaele_etal19, Fedorova_etal20}. This finding was further supported by numerical general circulation modeling \cite{Shaposhnikov_etal19,Neary_etal20}. It was suggested that this mechanism has likely governed the escape of water to space over geological time scales \cite{Fedorova_etal20}. The reported increase of GW activity at the very top of the atmosphere indicates that the waves not only contribute to the intensification of the transport, but can also directly boost the escape of hydrogen - a product of water photo-dissociation. The dominant process of its losses on Mars -- Jeans escape -- strongly depends on air temperature, which determines Maxwellian velocities of molecules.

Large density disturbances within the GW field imply similarly large variations of temperature: by 50 K on average and 100 K locally, based on relative density fluctuations and 250 K exobase mean temperature \cite{Medvedev_etal16}.
In order to illustrate the net increase of atmospheric losses induced by temperature variations associated with GWs, we consider the escape flux $\phi$ at the exobase. It is given by the expression \cite{Chaffin_etal18}
\begin{equation}
   \phi=n\frac{v_{mp}}{2\sqrt{\pi}} (1+\lambda)e^{-\lambda}, \quad  
    v_{mp}=\sqrt{\frac{2kT}{m}},  \quad \lambda=\frac{GMm}{kRT},
\end{equation}
where $n$ is the exobase density, $T$ is the exobase temperature, $v_{mp}$ is the most probable Maxwell-Boltzmann velocity, $\lambda$ is the escape or Jeans parameter, $k$ is the Boltzmann constant, $R$ is the exobase radius, $m$ is the mass of the hydrogen atom, $M$ is the planetary mass, and $G$ is the universal gravitational constant. The parameter $\lambda \approx 6$ at $T=250$ K at the Martian exobase \cite{Lammer2005}. The ratios of fluxes for wave-disturbed and undisturbed temperature $\frac{\phi(T+\delta T)}{\phi(T)}$ for sinusoidally varying temperature disturbance $\delta T$ are shown in Figure~\ref{fig:eflux} for two characteristic values: the reported 20\% (on average) and 40\% (locally). It is seen that the hydrogen escape flux increases by a factor of more than 2.5 and 5.5 at the peak of the positive phase for 20\%- and 40\% disturbances of temperature, respectively. The difference grows with the amplitude of fluctuations. Since the enhancement on the positive phase exceeds the reduction on the negative one, the net flux (integrated over the entire wave phase, the area shown with shades) also increases. For a 20\% and 40\% disturbances of temperature, the increase of the net escape flux is of 1.3 and 2, respectively. Note that this estimate does not account for wave-induced displacements of air parcels (pressure variations), which also contribute to the escape flux enhancement. 

Ordinarily, GW activity would be strongest when the thermosphere is coolest and vice versa, limiting escape as one effect canceled the other. However, dust storms reverse this paradigm, enabling larger wave amplitudes in a warmer background thermosphere. If the impact of the vertical water transport is considered, dust storms really represent a triple threat for atmospheric losses. Constraining the role of GWs in both transport and escape can thus help with quantifying the processes, which have made Mars a dry planet. 

%State-of-the art Martian general circulation models self-consistently coupling the different layers of the atmosphere should be used in the future to reveal the details of the dynamical interaction mechanisms between the lower and the upper atmosphere during global dust storms. 

\section{Summary and Conclusions}
\label{sec:concl}

Gravity wave-induced disturbances of CO$_2$ density obtained from the NGIMS instrument \changeE{onboard}{on board} MAVEN in the Martian thermosphere have been compared for two distinctive periods with the most close orbital coverage around the mid-year equinoxes: one during the dustless Martian Year (MY) 33 and the other in the midst of the MY34 global dust storm. For the first\removeE{,} time\addE{,} the net increase in Jeans escape due to GW-induced fluctuations \changeE{are}{is} estimated during the storm. The main results are listed below.
\begin{enumerate}
    \item GW activity approximately doubles during the dust storm. This estimate quantitatively agrees with that of \citeA{Leelavathi_etal20}, who considered Ar density fluctuations over a half-year period.
    \item The magnitude of relative density perturbations is around 20\% on average and 40\% locally. 
    \item The estimated net increase of Jeans escape flux of hydrogen is a factor of 1.3 and 2 for a 20\% and 40\% GW-induced disturbances of temperature, respectively.
\end{enumerate}

From a technological point of view, large GW-induced thermospheric density disturbances during dust storms can endanger spacecraft entries into the atmosphere, similar to aircraft that encounter bumpiness when flying over hills and mountains, and occasionally due to clear air turbulence. In all these cases, GWs are involved, and their forecasting is important and challenging.

\acknowledgments
The NGIMS level 2, version 8 data supporting this article are publicly available at 
%\newline
% \href{https://atmos.nmsu.edu/data_and_services/atmospheres_data/MAVEN/ngims.html}{\textcolor{blue}{https://atmos.nmsu.edu/data\_and\_services/atmospheres\_data/MAVEN/ngims.html}}
%\url{https://atmos.nmsu.edu/data_and_services/atmospheres_data/MAVEN/ngims.html}

\noindent
{\small \url{https://atmos.nmsu.edu/data_and_services/atmospheres_data/MAVEN/ngims.html}
}

%\bibliography{mybib-v51}

%%%%%%%%%%%%% FIGURES %%%%%%%%%%%%%%%%%
\clearpage

\begin{figure}
  \hspace*{-1.cm}
  \includegraphics[width=1.1\textwidth]{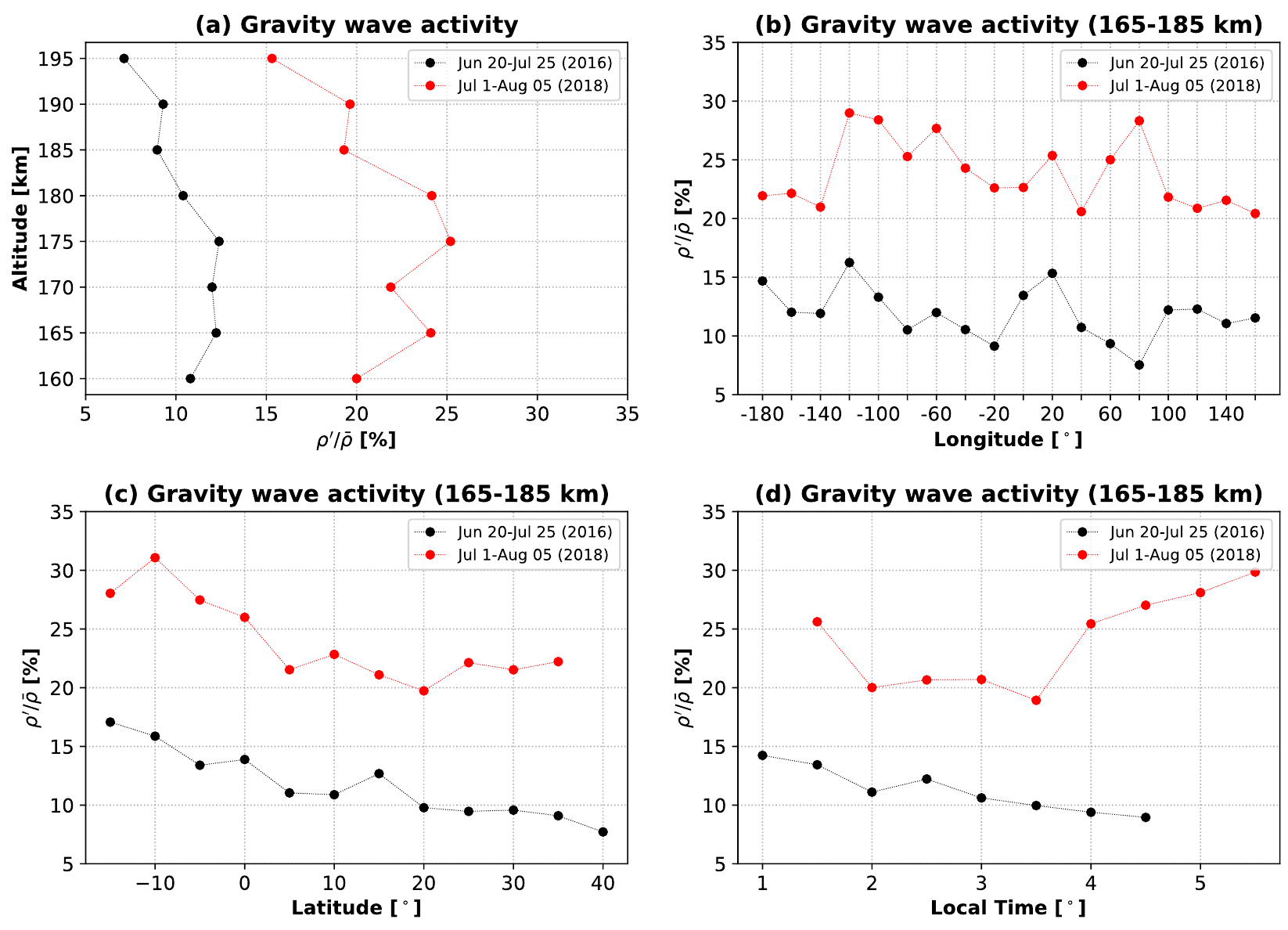}
  \caption{Comparison of gravity wave activity between the low-dust period in MY 33 $L_s = 171.7^\circ - 191.6^\circ$ (20 June -- 25 July 2016) and dust storm period in MY 34, $L_s=202.8^\circ-224.2^\circ$ (1 July -- 5 August 2018). (a) Altitude, (b) longitude, (c) latitude, and (d) local time variations of gravity wave activity under low dust conditions in 2016 and during the dust storm in 2018. The data is presented in terms of 5 km $20^\circ$, $5^\circ$, and 0.5 hour bins in (a)-(d), respectively.}
  \label{fig:2016vs2018}
\end{figure}

%% F3
\begin{figure}
  \vspace{-1cm}
 \hspace*{-1.5cm}  \includegraphics[width=1.2\columnwidth]{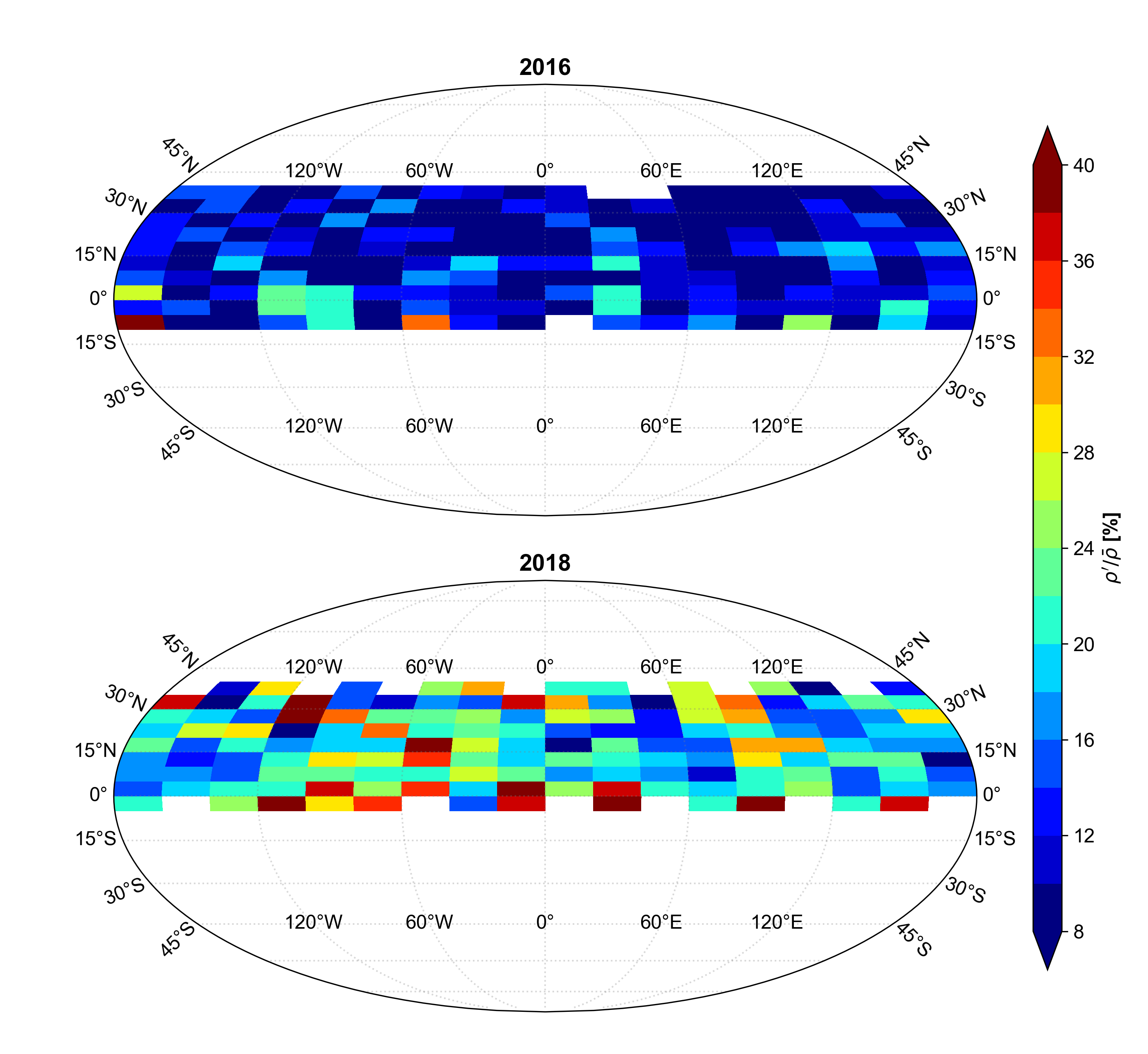}
 \caption{Comparison of the global distribution of the nighttime (1.5- 4.5 h) GW activity averaged within 165-- 185 km between the low-dust period in 2016 (MY 33, $L_s = 171.7^\circ-191.6^\circ$, 20 June-25 July 2016) and dust storm period in 2018 (MY 34, $L_s = 202.8^\circ-224.2^\circ$, 1 July-5 August 2018) presented in Fig 2. The data is binned in $20^\circ$, $5^\circ$ longitude-latitude bins.}
\label{fig:2016vs2018b}
\end{figure}

%% F4

\begin{figure}
  \vspace{-1cm}
 \hspace*{-1.5cm}  \includegraphics[width=1.25\textwidth]{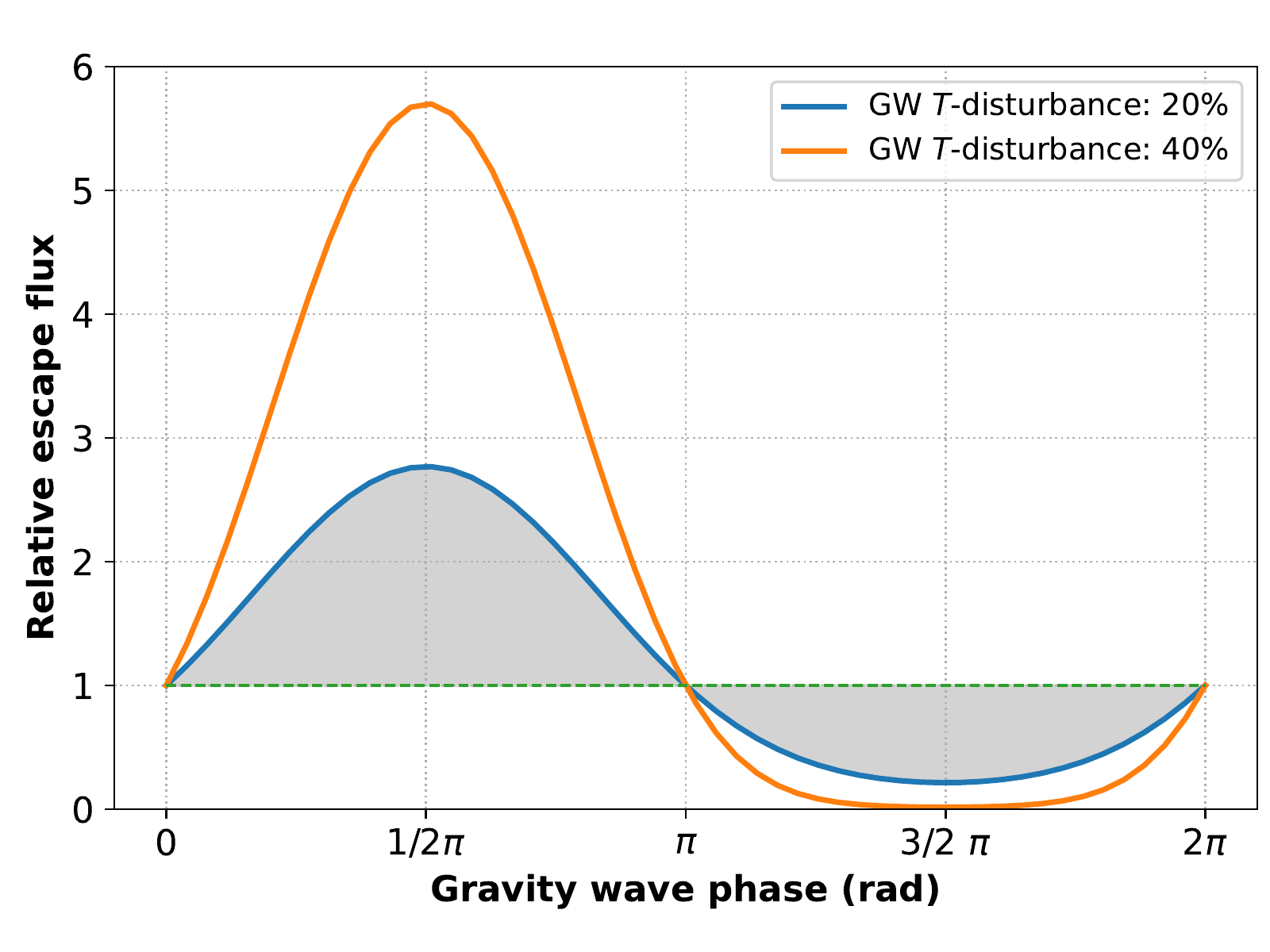}
 \caption{Relative escape flux $\frac{\phi(T+\delta T)}{\phi(T)}$ as a function of wave phase for the sinusoidally varying temperature disturbance $\delta T$. Blue and orange lines correspond to 20\% and 40\% amplitudes of fluctuations of the characteristic exobase temperature ($T_{exo}=250$ K) , correspondingly. The area under the curves gives the net (averaged over the entire wave phase) escape flux. Gray shading shows the net escape flux for 20\% amplitude of disturbances.}
\label{fig:eflux}
\end{figure}
\clearpage

\end{document}